\documentclass[10pt,a4paper,twoside]{article}
\usepackage{epsfig}
\usepackage{baltlat6}
\usepackage{array}
\begin{document}
\ \
\vspace{0.5mm}
\setcounter{page}{1}
\vspace{8mm}
 
\titlehead{Baltic Astronomy, vol.\,xx, xxx--xxx, xxxx}

\titleb{SOME PROPERTIES OF GALAXY STRUCTURES}
 
\begin{authorl}
\authorb{Piotr Flin}{1},
\authorb{Monika Biernacka}{1},
\authorb{W{\l}odzimierz God{\l}owski}{2},
\authorb{Elena Panko}{3} and
\authorb{Paulina Piwowarska}{2}
\end{authorl}
 
\begin{addressl}
\addressb{1}{Institute of Physics,  Jan Kochanowski University,\\
Swietokrzyska  15,  25-406  Kielce, Poland, sfflin@cyf-kr.edu.pl,bmonika@ujk.edu.pl}
\addressb{2}{Institute of Physics,  Opole University,\\
Oleska 48, 45-041  Opole,Poland, godlowski@uni.opole.pl,paoletta@interia.pl;}
\addressb{3}{Kalinenkov Astronomical Observatory, Nikolaev State University,\\
Nikolskaya 24,  Nikolaev, 54030 Ukraine,  panko.elena@gmail.com;}
\end{addressl}
 
\submitb{Received: ; accepted: }
 
\begin{summary}
 
We analysed some properties of galaxies structures based on PF Catalogue
of galaxy structures (Panko \& Flin 2006) and Tully NBG Catalog (Tully 1988)
At first, we analyzed the orientation of galaxies in the 247 optically selected
rich Abell clusters, having in the area considered as cluster at least
100 members. The distribution of the position angles of the galaxy as well as
two angles describing the spatial orientation of galaxy plane were tested for
isotropy, applying three statistical tests. We found that anisotropy
increases with the number of member galaxies, which means there exists the
relation between anisotropy  and cluster richness. We do not find connection
of galaxy alignment and Bautz - Morgan morphological type of parent cluster.
The statistically marginal relation between  velocity dispersion and cluster
richness was observed. However it was found that velocity dispersion decreases
with  Bautz - Morgan type at almost $3 \sigma$ level. Separately we analyzed 
ellipticities for 6188 low redshift ($z < 0.18$) poor and rich galaxy 
structures which  have been examined along with their evolution. Finally we 
analysed the Binggeli effect. The orientation of galaxy groups in the Local 
Supercluster (LSC), is strongly correlated with the distribution of 
neighbouring groups in  the scale till about 20 Mpc. During analyses of galaxy 
structures from PF catalogue the situation was quite different. The efect is
observed only for more elongated structures ($e \le 0.3$). The range of the
distance in which the effect was observed, was estimated as about
$60 h^{-1} Mpc$.
 
\end{summary}

 
\resthead{Some properties of galaxy structures}
{P. Flin, M. Biernacka, W. God{\l}owski, E. Panko, P. Piwowarska}

\sectionb{1}{INTRODUCTION}
 
The formation of large scale structures is one of the most important problem
in the modern astrophysics. There are many theories called scenarios of
structure formation (Peebles 1969, Zeldovich 1970, Sunyaew \& Zeldovich 1972,
Doroshkevich 1973, Shandarin 1974, Dekel 1985, Wesson 1982,
Silk \& Efstathiou 1983, Bower et al. 2005). In the presently most popular
$\Lambda$CDM model, the structures were formed from the primordial
adiabatic, nearly scale invariant Gaussian random fluctuations
(Silk 1968, Peebles \& Yu 1970, Sunyaew \& Zeldovich 1970). A very important
problem is to  distinguish between different models of galaxy and galaxy
structure formation. An investigation of the orientation of galaxies in
clusters is regarded as a standard test of theories of galaxy and large scale
structure formation. Thus, theories of the galaxy formation make predictions
regarding to the angular momenta of galaxies (Peebles 1969, Doroshkevich 1973,
Shandarin1974, Silk 1983, Catelan \& Theuns 1996, Li 1998, Lee \& Pen 2002,
Trujillio, Carretero and 2006). Because this  parameter is known  for small
number of galaxies, instead of the angular momenta, the orientation of member
galaxies in structure is studied. There are two approaches to this problem,
because either the distribution of galaxy position angles or the distribution
of the orientation of galaxy planes can be investigated. The analysis of the
orientation of  galaxies is performed on the sample of 247 optically selected
rich Abell clusters, having in the area considered as cluster at least 100
members. The distribution of the orientation of the brightest galaxies in
clusters as well as an analysis of the cluster elipcities is presented for
a statistically complete sample of 6188  galaxy structures from PF Catalogue
(Panko \& Flin 2006 hereafter PF). Moreover we analysed the orientation of
galaxy groups in the Local Supercluster and alignment of galaxy clusters.

\sectionb{2}{OBSERVATIONAL DATA}

The PF Catalogue of galaxy structures (Panko \& Flin 2006) is the observational
basis for the present study. The catalogue was created using data
from the Muenster Red Sky Survey (Ungruhe, Seitter, Duerbeck 2003, hereafter
MRSS), which is a large-scale galaxy catalogue covering an area
of about 5000 square degrees in the southern hemisphere,
complete to a magnitude limit of $m =18.^m 3$. The same magnitude
limit defines the completeness limit for galaxies in the PF Catalogue.
The MRSS is the result of scanning 217 ESO plates with $b < -45^o$.
 
The 2D Voronoi tessellation technique (Ramella et al. 2001, 
Panko and Flin 2006) was applied to the galaxy catalogue
to search for overdense regions. As a result PF includes
6188 such structures, with at least 10 galaxies in each
structure field. The Voronoi procedure gives only the area and
equivalent radius for the overdense structures, while the PF
contains information about their shape and orientation on the
celestial sphere. A covariance ellipse method involving five
moments for the distribution of galaxy coordinates was used to
calculate the elliptical shape describing each structure as
well as the position angle of its long axis. Lists of galaxies
in the magnitude range $m_3$ to $m_3+3^m$, where $m_3$ is the brightness
of the third brightest galaxy in the structure area, and the
rectangular coordinates x and y of the galaxies were employed
for the calculation of the semiaxes a and b of the resulting
ellipses, the ellipticity parameter , and the position angle of
each structure's long axis.
 
Since the MRSS does not contain galaxy distances, in order to
obtain distance estimates for the PF structures, we calibrated
the $(log z)-m_{10}$ relation following Dalton et al. (1997). The
first step of the procedure was to compare the positions of the
structure centers, as given in the PF Catalogue, with those in the
ACO ( Abell et all.  1989) and APM cluster (Dalton et al. 1997)
If the distance between the centers of the PF and ACO clusters
was less than 0.5 of the PF  equivalent cluster radius, the two
objects were regarded as
identical. More than 1000 such identifications were found. Only
466 ACO clusters from the list have measured redshifts z in
NED. The calibration of the $(log z)-m_{10}$ relation in the form:
$log z_{est}= a + b*m_{10}$   based on 455 data points is illustrated
in Fig. 1 of papers Biernacka, Flin and Panko (2009) and Panko et al. (2009a).
The structure identification was repeated using a value of 0.3 of the PF
cluster radius as a new criterion of identity. That produced a
similar relationship, but based on a smaller number of center
coincidences.
 
  We  repeated the procedure with data from APM clusters, taking
the  measured redshift $z$ from the APM cluster catalogue (Dalton
et  al.  1997). Additional calibration of the $log  z_est - m_{10}$
relation was possible through the comparison of a deeper version of
the PF catalogue with the ACO catalogue. The deeper version  of
the  catalogue  is  not  statistically  complete,  because  the
limiting  magnitude of considered galaxies is $r_F=19^m.3$.   For
future  work  we  select the relation: $log z_{est}=-3.771+0.1660*m_{10}$.
Each particular relation is located  within  the
confidence  limits for the other relations: but even  the  most
discordant relations give a maximal difference in $z_est$ of only
0.02 for a magnitude limit of $m=18^m.3$.

\sectionb{3}{THE METHOD OF INVESTIGATION }

\begin{figure}
\vskip 4cm
\includegraphics{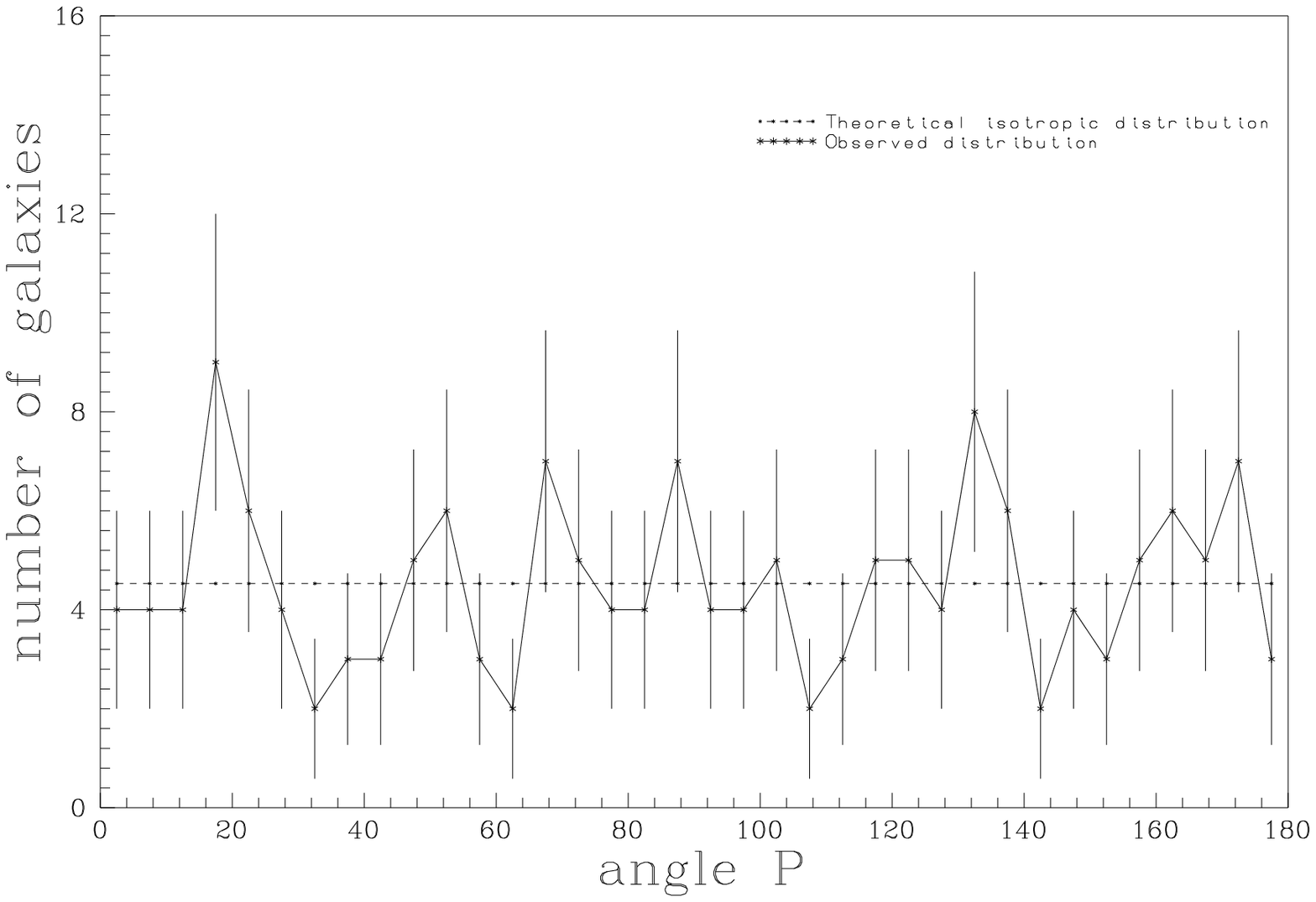}
\includegraphics{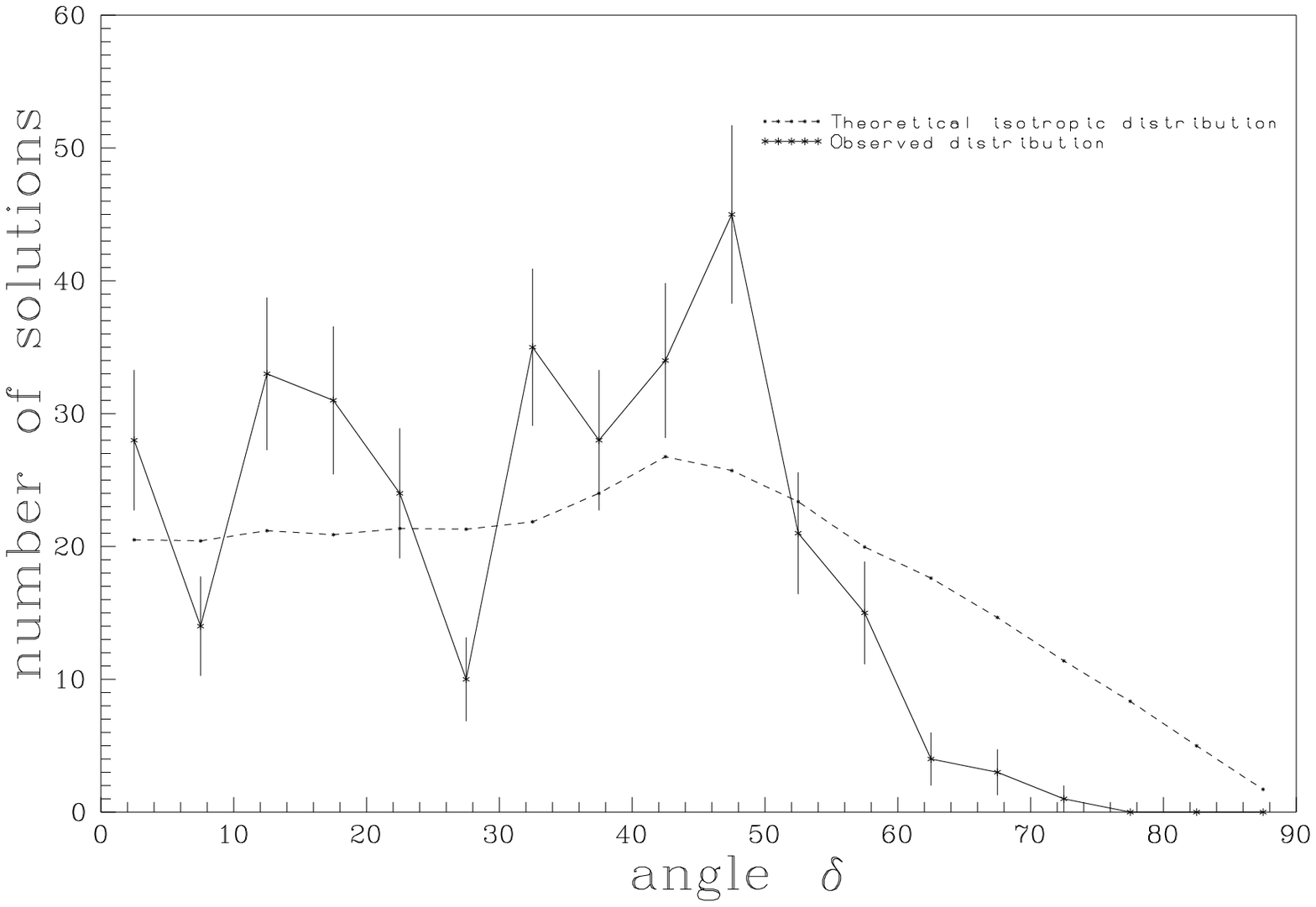}
\includegraphics{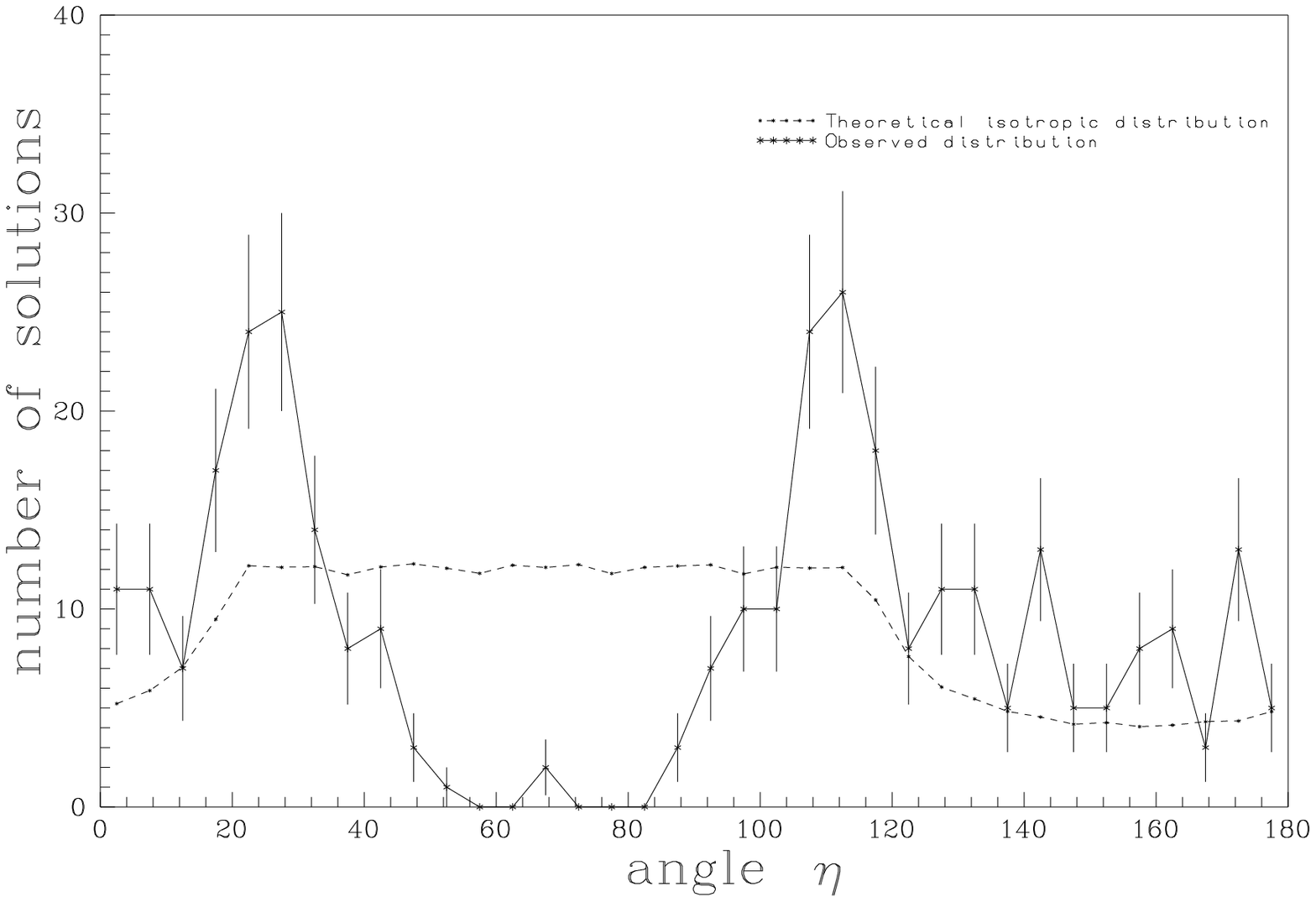}
\includegraphics{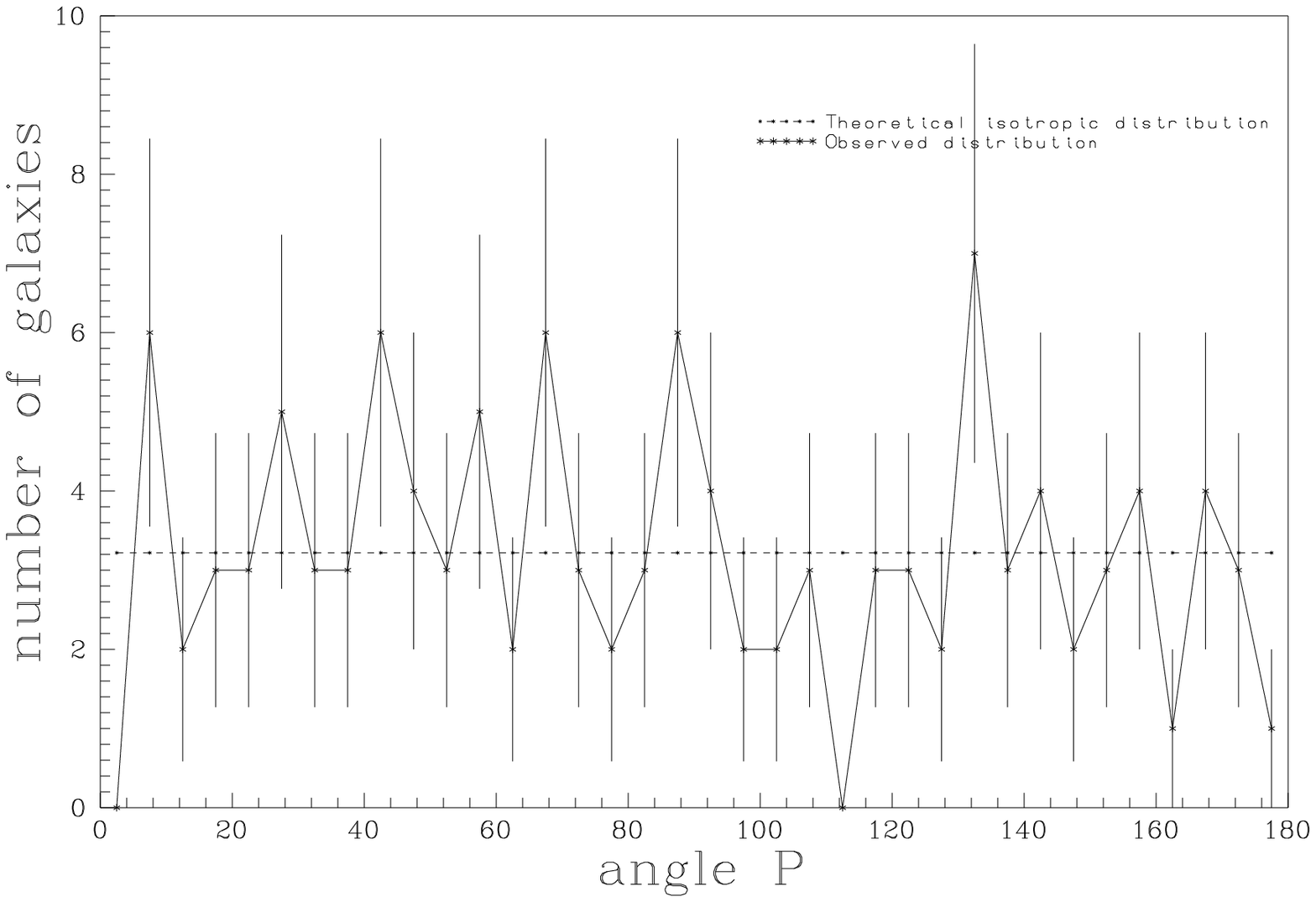}
\includegraphics{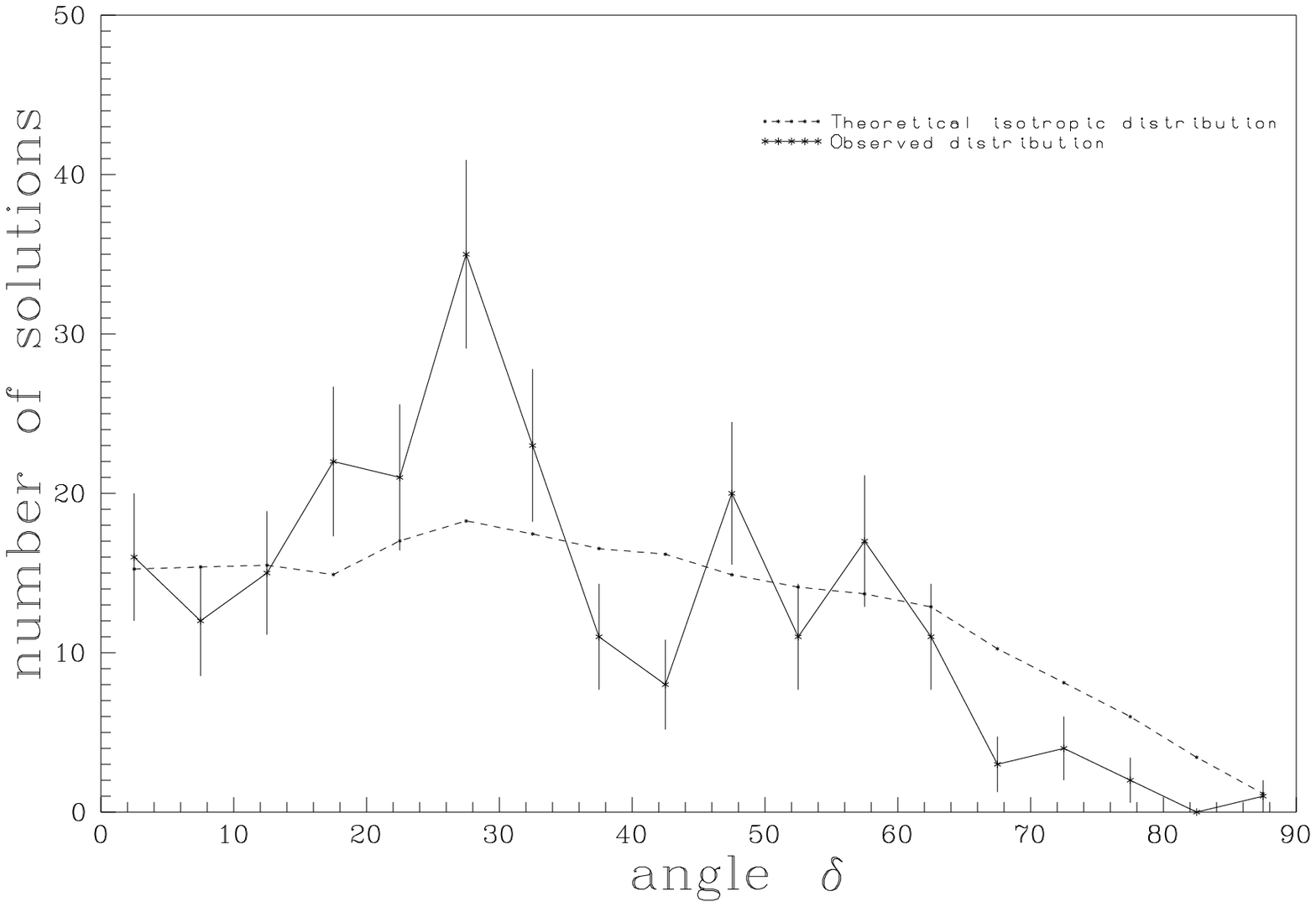}
\includegraphics{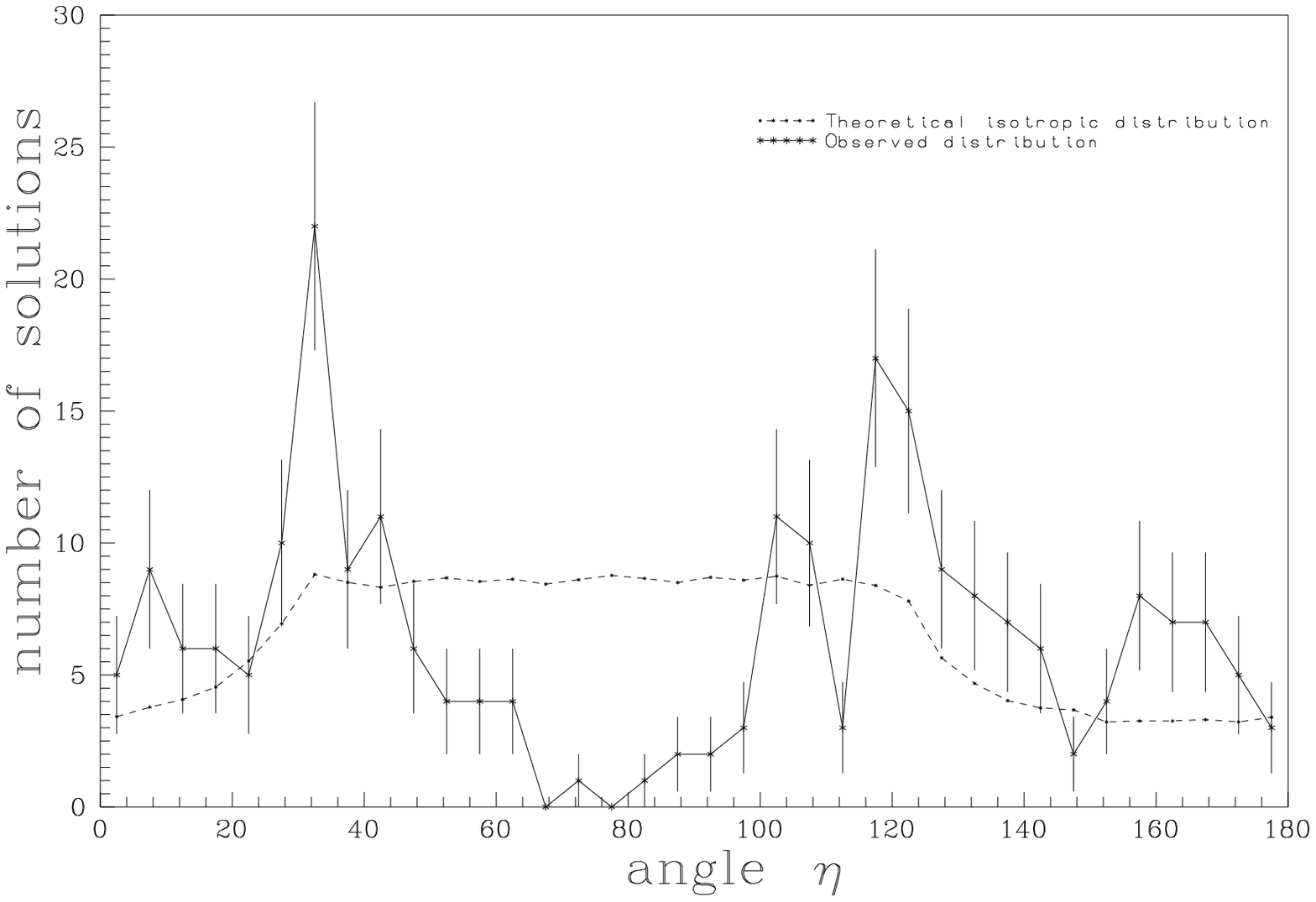}
\caption{The  distribution of the position angels - left panel, $\delta_D$
angles - middle panel and $\eta$ angles - right panel, for clusters: 
A2721 - upper panel and - A2554 bottom panel (supergalactic coordinate system,
galaxies with $b/a \le 0.75$). Theoretical distribution (dashed line) and the
error bar were presented as well.}
\end{figure}

There are two methods for  galaxy orientation  study.  In the first one
(Hawley \& Peebles 1975) the analysis of the distribution of the observed
position angle of the galactic image major axis was carried out.
The second approach allowed us to use also the face-on galaxies. This method,
based on the de-projection of the galaxy images, was originally proposed by
{\"O}pik (1970),  applied by Jaaniste \& Saar (1977) and significantly
modified by Flin \& God{\l}owski (1986, 1989) and God{\l}owski (1993, 1994).
In this method, the  galaxy's inclination with respect to the observer's line
of sight $i$ is  considered. Two possible orientations of the galaxy plane
were determined, which gave two possible directions perpendicular to the galaxy
plane. It is expected that one of these normals corresponds to the direction
of galactic rotation axis. Such study gives a four-fold ambiguity in the
solution for angular momentum. By the reason of no information connected with
 the direction of the galaxy spin our analysis is reduced to only two 
 solutions. The inclination angle was calculated according to the formula
valid for oblate spheroids Holmberg (1946):
$cos^2 i=(q^2 -q^2_0 )/(1-q^2_0)$, where observed axial ratio $q=b/a$ and
$q_0$  is  "true" axial ratio.  We  used standard value $q_0=0.2$.
For each galaxy, two angles are  determined: $\delta_D$ - the
angle between the normal to the galaxy plane and the main plane of the
coordinate system and $\eta$ - the angle between the projection of this normal
onto the main plane and the direction towards the zero initial meridian.
Using the equatorial coordinate system, the following relations hold between
angles ($\alpha$, $\delta$, $p$) and ($\delta_D$, $\eta$)
\begin{equation}
\sin\delta_D  =  -\cos{i}\sin{\delta} \pm \sin{i}\cos{r}\cos{\delta},
\end{equation}
\begin{equation}
\sin\eta  =  (\cos\delta_D)^{-1}[-\cos{i}\cos{\delta}\sin{\alpha} + \sin{i}
(\mp \cos{r}\sin{\delta}\sin{\alpha} \pm \sin{r}\cos{\alpha})],
\end{equation}
where $r=p-\pi/2$.
In order to detect non-random effects in the distribution of the investigated
angles: $\delta_D$, $\eta$ and $p$ we carried out three different statistical
tests. These test were : the $\chi^2$  test, the autocorrelation test and the
Fourier test  (Hawley \& Peebles 1975, God{\l}owski 1993, 1994).
Studying  the distribution of  position
angles, face-on and nearly face-on galaxies were excluded from the analysis.
In this case  only  galaxies  with axial ratio $b/a \le 0.75$  were taken into
consideration.  In our previous paper (God{\l}owski et al. 2010),
during analysis of $\delta_D$ and $\eta$ angles, we took into account all
galaxies, including face-on. In the present paper also in this case we taking
into account only galaxies with axial ratio $b/a \le 0.75$. It mean that we
exclude face-on and nearly face-on galaxies. We checked whether the
distributions of the investigated angles ($\delta_D$, $\eta$ and $p$) in each
individual cluster  were isotropic (for details see God{\l}owski, et. al 2010).

\sectionb{4}{RESULTS}
 
\subsectionb{4.1}{The dependence of  alignment on the cluster richness}

\begin{figure}
\vskip 4.8cm
\includegraphics{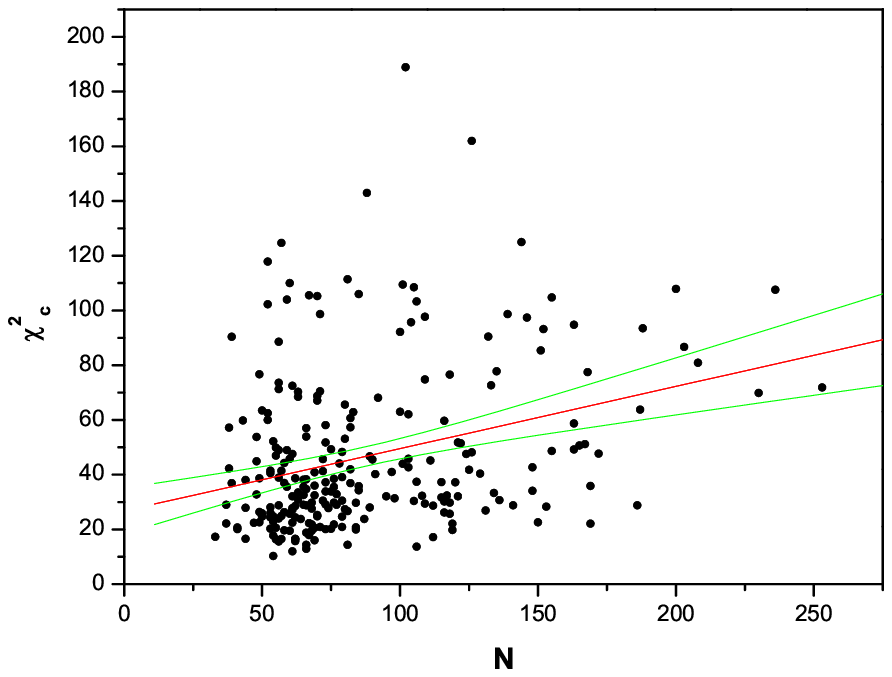}
\includegraphics{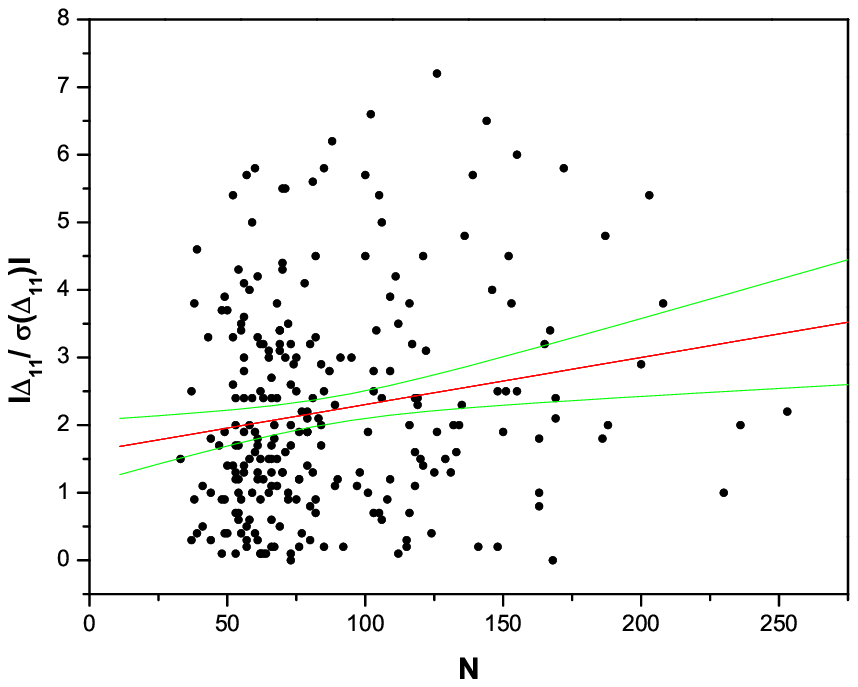}
\includegraphics{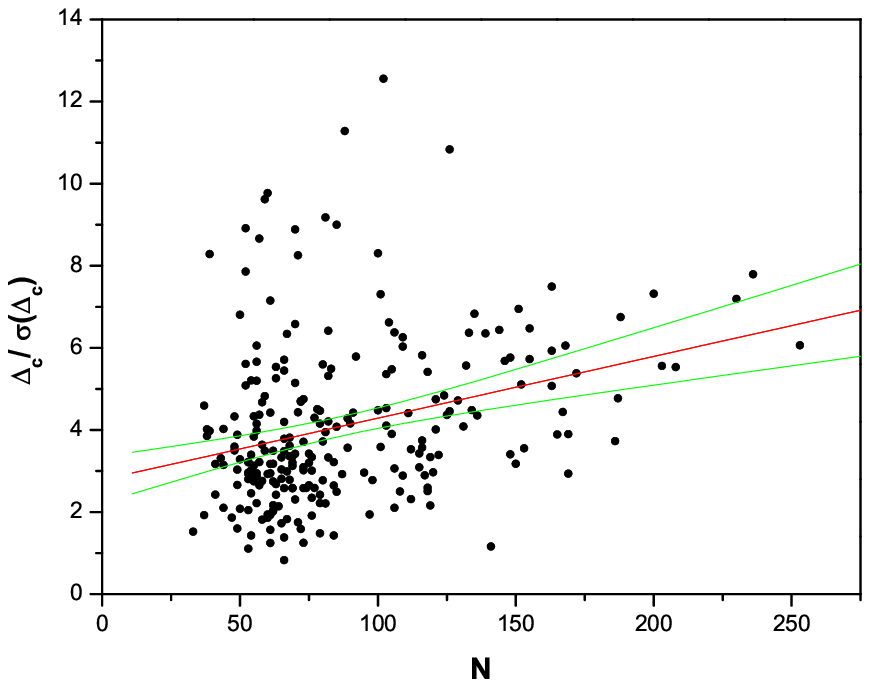}
\includegraphics{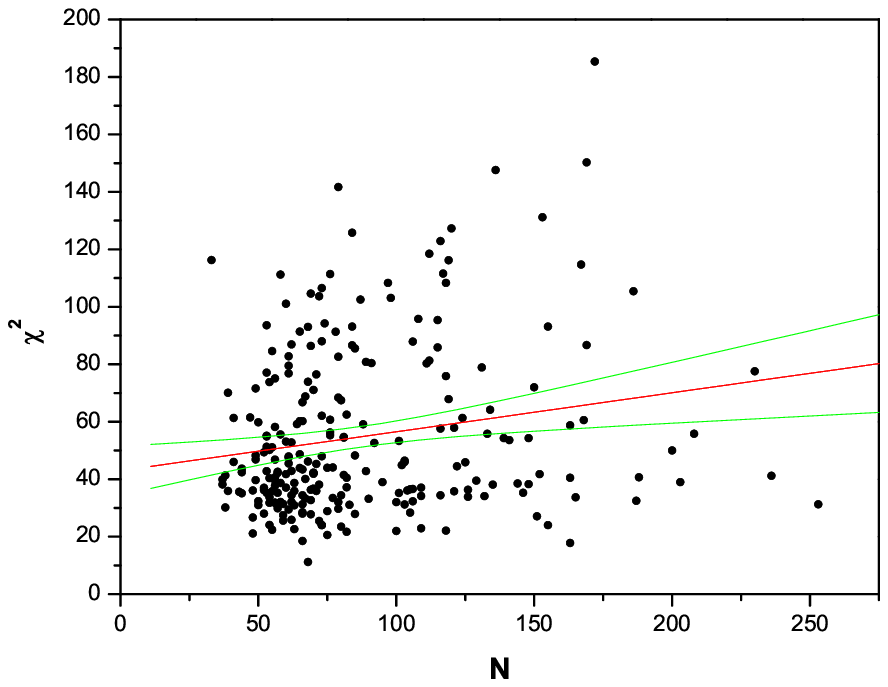}
\includegraphics{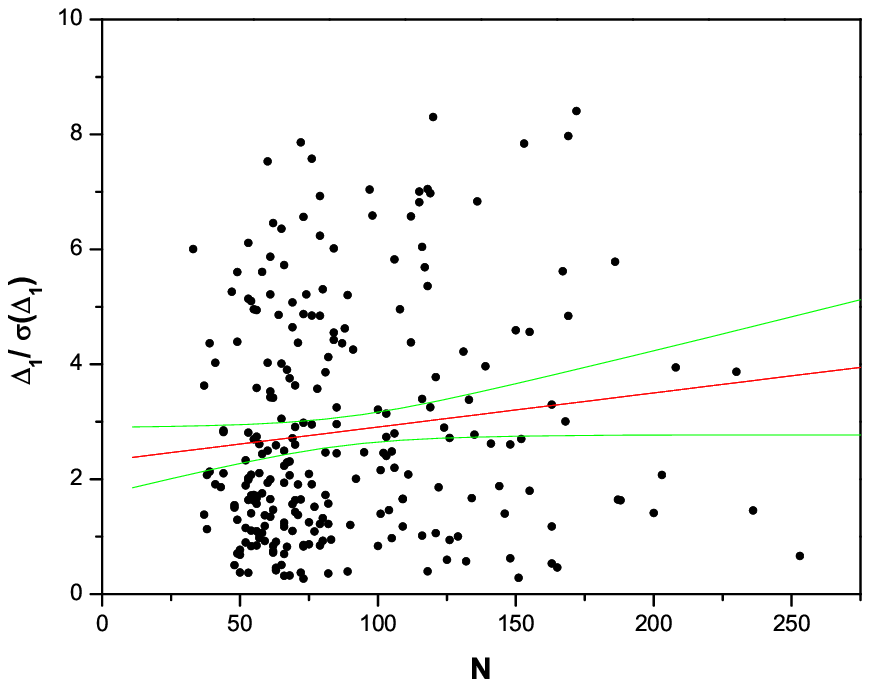}
\includegraphics{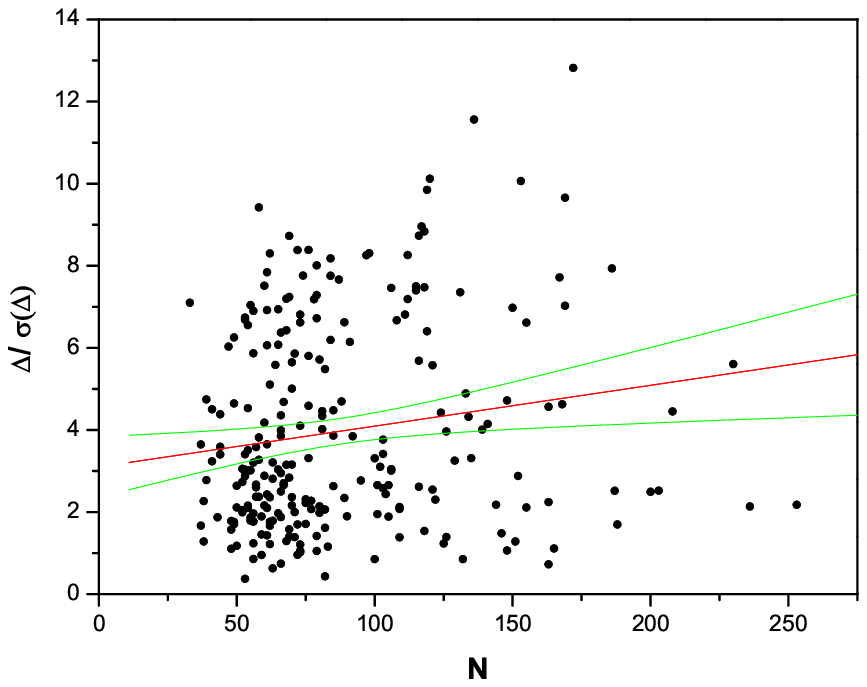}
\caption{The relation  between the number of galaxies in the cluster $N$
and the value of analyzed statistics ($\chi^2$ - left panel,
$\Delta_1/\sigma(\Delta_1)$ - middle panel, $\Delta/\sigma(\Delta)$ -
right panel)  for the analysed angles - equatorial coordinates. Upper panel -
$\delta_D$ angles, bottom panel - $\eta$ angles. The bounds error, at
confidence level $95\%$, were presented as well (galaxies with 
$b/a \le 0.75$).}
\end{figure}

Expected isotropic distribution of $\delta_D$ and $\eta$ (polar and azimuthal)
angles are modified if some galaxies (for example face on galaxies) are
excluded from the analysis. In such situation expected isotropic distribution
should be found from Monte Carlo simulation. In the paper
God{\l}owski et al. (2010) during analysis of $\delta_D$ and $\eta$ angles
we took into account all galaxies. Figure 1 in this paper shows
the observed and theoretical distribution the investigated angles
($\delta_D$, $\eta$ and $P$) for two studies in the past, cluster A2254 and
A2771. Now we present appropriate theoretical and observed distribution of
$P$, $\delta_D$, and $\eta$ angles for clusters A2254 and A2771 (Fig.1) in
the case when we excluded galaxies with $b/a >0.75$. Please note that our
plots are very similar to that obtained by Aryal, Paudel, Saurer (2007).

On the base of our previous papers (God{\l}owski, Baier and Mac Gillivray 1998,
God{\l}owski \& Ostrowski 1999, Baier, God{\l}owski and  Mac Gillivray 2003
and God{\l}owski, Szyd{\l}owski Flin 2005) the suggestion was found that
we do not observe any  alignment in sparsely populated clusters, while  such
alignment  is observed in a number of rich clusters of  galaxies. However,
this suggestion (God{\l}owski, Szyd{\l}owski and Flin 2005 and later confirmed
by Aryal,  Pudel  and Saurer 2007)  was only a qualitative  and  
quantitative analysis on this issue was not provided.

 \begin{table}
\begin{center}
\scriptsize
\caption{The results of the linear regression analysis - equatorial coordinates.
\label{Tab.1}}
\begin{tabular}{c|cccc|cccc|cccc}
\hline
\multicolumn{1}{c}{}&
\multicolumn{4}{c}{$\chi^2$}&
\multicolumn{4}{c}{$\Delta_1/\sigma(\Delta_1)$}&
\multicolumn{4}{c}{$\Delta/\sigma(\Delta)$}\\
\hline
sample&$a$&$\sigma(a)$&$b$&$\sigma(b)$&$a$&$\sigma(a)$&$b$&$\sigma(b)$&$a$&$\sigma(a)$&$b$&$\sigma(b)$\\
\hline
$P$     &$0.025$&$ 0.015$&$34.7$&$1.4$&$0.0018$&$0.0015$&$1.55$&$0.15$&$0.0022$&$0.0015$&$2.08$&$0.14$\\
$\delta$&$0.227$&$ 0.044$&$26.8$&$4.3$&$0.0069$&$0.0024$&$1.61$&$0.23$&$0.0015$&$0.0028$&$2.78$&$0.29$\\
$\eta$  &$0.135$&$ 0.045$&$43.0$&$4.3$&$0.0059$&$0.0031$&$2.31$&$0.30$&$0.0099$&$0.0039$&$3.10$&$0.38$\\
\hline
\end{tabular}
\end{center}
\end{table}
 
    Therefore God{\l}owski et al. (2010) examined orientation  of
galaxies  in  clusters both qualitatively and quantitatively  and found
a  sharp  increase  of galaxy orientation  alignment  with richness
of the cluster. Both in the present paper and God{\l}owski et al. (2010)
we analyzed the orientation of  galaxies in the 247 optically
selected very  rich Abell clusters, having in the area
considered as cluster at least 100 members. In the paper 
God{\l}owski at al. (2010) during analysis of $\delta_D$ and $\eta$ angles
we took into account all galaxies (including face-on ones).
Now we present the analysis of $\delta_D$, $\eta$ angles in the case when
galaxies seen face-on and nearly face-on are excluded.
The position angles of the galaxy major axis image  as well as two angles 
$\delta_D$, and $\eta$  describing the spatial orientation of galaxy plane
were tested for isotropy, applying three different statistical tests.
We investigated  the relation between the values of the applied statistics 
and the cluster richness for investigated angles, both in equatorial
and supergalactic coordinate system.  The results are shown on
Fig.2  and Table 1. We found that the values of applied statistics increase
with the number of member galaxies, which means  the relation between
anisotropy  and cluster  richness. The search for connection of galaxy
alignment and Bautz - Morgan morphological type of parent cluster gave only
weak dependence. The statistically marginal relation between  velocity
dispersion and cluster richness was observed.  At almost $3 \sigma$ level
velocity dispersion decreases with  Bautz - Morgan type.
The effect increases if we restricted the cluster membership to galaxies
brighter than $m_3+3$, which suggests that this effect is really connected
with clusters. We also repeated the analysis using sample of galaxies
containing magnitude range $m_3+2$. There was no significant differences
between results of linear regression for sample with magnitude range $m_3+2$
and $m_3+3$. One should note that excluding from analysis face-on and nearly
face-on galaxies does not change our results significantly with comparision
to the results God{\l}owski et al. (2010) paper.
 
These results showing the  dependence of galaxy  alignment on either the
cluster richness or other tested parameters are due to  environmental  effect.
In our opinion the observed relation between the richness of galaxy cluster and
the alignment is due to tidal torque, as suggested by Catelan \& Theuns 1996).
One should note however, that our finding is also in agreement  with prediction
of the Li model (Li 1998, God{\l}owski  Szyd{\l}owski and Flin 2003).

\subsectionb{4.2}{The alignment of the brightest cluster galaxy with the
parent  cluster}
 
For 1056 PF structures, which has been identified with ACO 
(Abell et al. 1989) clusters, with known BM morphological types we
studied  the alignment of the brightest galaxy with the parent
cluster.  We investigated  for isotropy acute  angle $\phi$
being the difference  between the structure position angle 
and the position angle of the brightest cluster galaxy  (Panko
et al. 2009b).  Several subsamples were analaysed for isotropy
of the  $\phi$ angle.  The distance between clusters as well as
the BM morphological types served for subsamples selection.
The isotropy  of the distrbtion was  tested by applying
statistcal  tests mentioned  above (section 4.1).  Only in the
case  of  clusters  BM I  the distribution was non random; the
small excess of small $\phi$ angles was observed, which means that
the brightest  cluster member tends  to be aligned with the
parent  cluster.  In Fig.3 two distributions of the  $\phi$
angles for cluster of BM types I and I-II are given. (Panko et al. 2009b).
Moreover the orientation of the tenth brightest galaxies  in
each   structure   was  checked  for  the   isotropy.   The
distribution of $PA_i$   (where i=1,2,3,10, ) appeared to  be
isotropic (Panko et al. 2009b). A separate test was performed
to establish whether the observed angle in a particular bin
deviates from a isotropy by more than  a  standard
deviation $\sigma = \sqrt{N}$ - where N is the number  of  objects
expected in the bin for a random distribution.
Separately, the isotropy  of the distribution of 6188 structures
position angles, as well as 1056 PF structures identified  with
ACO clusters was  tested.
In any cases, we do not find any significance deviation from isotropy.
Therefore, we concluded, that the  analysed distributions were isotropic.
 



\begin{figure}
\vskip 3.4cm
\includegraphics{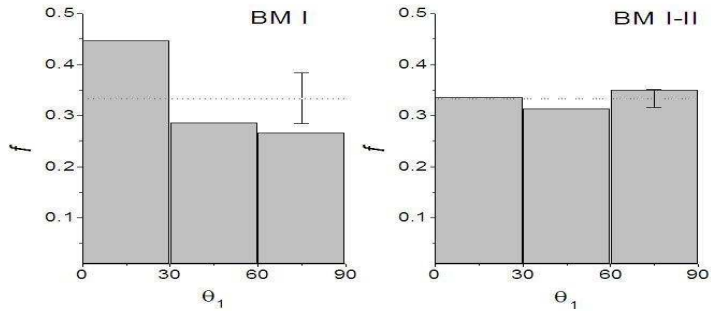}
\caption
{The frequency distribution of the acute angle between the parent cluster
and the brightest galaxy positional angles for BM type I and BM I-II. Dotted
lines show the isotropic distribution together with $1\sigma$ error.}
\end{figure}

\subsectionb{4.3}{The alignment  of galaxy clusters}
 
Binggeli (1982) found that galaxy clusters tend to  point each
other.  We search for Binggelli's efect through invesigations
of the dfference  between the postion angle  of cluster and
direction toward neigbouring structures. The resulting acute
angle $\phi$  was tested for isotropy. Only after restrictng our samples to
more elongated structures  ( $e \le 0.3$) the distributions   became
anisotropic, with the excess of small   $\phi$  angles, which
indcates the existece of the Binggeli's effect. This effect was
statistically weak and observed  in the case of all 1056
structures, as well as after dividing according to BM types. The
strongest effect was for BM I and BM III. The  range of the
distance  in which the effect was observed, was about $60 h^{-1} Mpc$.

\subsectionb{4.4}{The orientation of galaxy groups in the Local Supercluster}
 
We analysed also the orientation of galaxy groups in the Local Supercluster
(God{\l}owski \& Flin 2010).
It is strongly correlated with the distribution of neighbouring groups in the
scale till about 20 Mpc.  The group major axis is in alignment with both
the line joining the two brightest galaxies  and the direction toward the
centre of the LSC. These correlations suggest that two
brightest galaxies were formed in filaments of matter directed towards the
protosupercluster centre. Afterwards, the hierarchical clustering leads to
aggregation of galaxies around these two galaxies. The groups are formed on
the same or similarly oriented filaments.

\sectionb{5}{CONCLUSIONS}
 
We study the distribution of several parameters connected with the
angular momenta of galaxies. In the $\Lambda$CDM model, which is the most
popular one, the angular momenta of galaxies are mainly due to tidal torquing
of neigbouring galaxies. Our result that alignment of  galaxies in rich cluster
is increasing with the number of galaxies in clusters is due to tidal torque, 
as suggested by Catelan \& Theuns (1996). One should note however, that our 
finding is also in agreement  with prediction of the Li model (see also 
God{\l}owski 2011).
 
The  observed alignment of the brightest cluster galaxy with the
parent cluster for   BM I type clusters  shows the special role played by
gigantic cD galaxy during the evolution of cluster. The alignment of groups 
in the LSC is observed till  about $20h^{-1}$ Mpc.  The  group major axis is 
aligned with both the line
joining the two brightest galaxies and the direction  toward the centre of
the LSC, this is Virgo Cluster. Two brightest galaxies were formed on the
filament of matter directed toward the protocluster centre. Afterwards,
the hierarchical clustering leads to aggregation of galaxies  around these two
galaxies. The groups were formed on the same or similarly oriented filaments.
This picture is in agreement with both the prediction of numerical simulation
and  expectation of $\Lambda$CDM models.
 
The covariance ellipse method is a good tool for cluster ellipticity studies 
(Biernacka \& Flin 2011). The shape of  each structure projected on the
celestial sphere was determined using the covariance ellipse method. Analysis
of the data indicates that structure ellipticity changes  with redshift.
Nearer structures are more round, which can be  attributed to virialisation
 processes occurring in the not so distant past (see also Struble \&. Ftaclas)

\thanks{The present research has made use of the NASA/IPAC Extragalactic
Database (NED), which is operated by the Jet Propulsion Laboratory, California
Institute of Technology, under contract with the National Aeronautics and Space.
This research has also made use of NASA's Astrophysics Data System. The present
research was partially supported by the Jan Kochanowski University grants BS 052
and BW116. EP thanks the Jan Kochanowski University for support during her stays
in Poland.}
 
\References
 
\refb Abell, G.O., Corin, H.G., Olowin, R.P., 1989, APJS, 70,1

\refb Aryal, B., Paudel, S., Saurer, W. 2007, MNRAS, 379, 1011
 
\refb Baier, F., God{\l}owski, W., Mac Gillivray H.T., 2003, A\&A, 403, 847
 
\refb Biernacka, M., Flin, P., Panko, E., 2009, ApJ

\refb Biernacka, M., Flin, P.,  2011, AN, 332, 537
 
\refb Binggeli, B., 1982, A\&A, 107, 338
 
\refb Bower, R. G., Benson, A.J., Malbon, R., et al. 2006, MNRAS, 370, 645
 
\refb Catelan, P., Theuns, T. 1996, MNRAS, 282, 436
 
\refb Dalton, G., Maddox, S., Sutherland W., Efstathiou, G., 1997. MNRAS 289, 263

\refb Dekel, A., 1985, APJ, 298, 461
 
\refb Doroshkevich, A. G. 1973, APJL, 14, 11
 
\refb Flin, P., God{\l}owski, W., 1986, MNRAS, 222, 525
 
\refb Flin, P., God{\l}owski, W., 1989 Sov.Astron.Lett. 15, 374 (Pisma Astr.Zur. 15, 869)
 
\refb God{\l}owski, W. 1993, MNRAS, 265, 874
 
\refb God{\l}owski, W. 1994, MNRAS, 271, 19
 
\refb God{\l}owski, W. 2011, IJMPD (accepted) arXiv 1103.5786
 
\refb God{\l}owski, W., Baier, F., Mac Gillivray, H.T., 1998, A\&A, 339, 709
 
\refb God{\l}owski, W., Ostrowski, M., 1999, MNRAS, 303, 50
 
\refb God{\l}owski, W., Flin, P., 2010 APJ, 708, 920
 
\refb God{\l}owski, W., Piwowarska, P., Panko, E., Flin, P., 2010,  APJ, 723, 985
 
\refb God{\l}owski, W., Szyd{\l}owski, M., Flin, P., Biernacka, M., 2003 GRG, 35, (5), 907
 
\refb God{\l}owski, W., Szyd{\l}owski, M., Flin, P. 2005 GRG, 37, (3) 615
 
\refb Hawley, D. I., Peebles P. J. E. 1975, AJ, 80, 477
 
\refb Holmberg, E. 1946, Medd. Lund. Astron. Obs. Ser. VI, Nr. 117

\refb Jaaniste, J. Sarr, E. 1977 Tartu Obs. Preprint A-2
 
\refb Lee, J., Pen, U. 2002, APJ, 567, L111
 
\refb Li, Li-Xin. 1998, GRG, 30, 497
 
\refb {\"O}pik, E.J. 1970, Irish AJ, 9, 211

\refb Panko, E., Flin P., 2006 The Journal of Astronomical Data 12, 1 

\refb Panko,  E., Juszczyk, T., Biernacka, M.,  Flin, P.,  2009a, APJ, 700, 1686

\refb Panko,  E., Juszczyk, T., Flin, P.,  2009b, AJ, 138, 1709

\refb Peebles, P.J.E. 1969, APJ, 155, 393
 
\refb Peebles, P.J.E., Yu, J., T. 1970, APJ, 162, 815
 
\refb Ramella M., Boschin W., Fadda D., Nonino M. 2001,  A\&A, 368, 776
 
\refb Silk, J. 1968, APJ, 151, 459
 
\refb Silk, J., Efstathiou, G. A. 1983,  Fundamentals of Cosm. Phys. 9, 1

\refb Struble, M., F. Ftaclas C. 1994, AJ 108, 1 

\refb Sunyaev, A. R., Zeldovich, Ya. B., 1970, Astroph. Sp. Sci., 7, 3
 
\refb Sunyaev, A. R., Zeldovich, Ya. B., 1972 A\&A, 20, 189
 
\refb Shandarin, S.F. 1974, Sov. Astr. 18, 392
 
\refb Trujillo, I., Carretero C., Patri G. 2006, APJ, 640, L111
 
\refb Tully, R. B. 1988, Nearby Galaxy Catalog, Cambridge Univ.Press: Cambridge
 
\refb Ungruhe, R., Seitter, W., Durbeck, H.  2003, Journal of Astronomical Data, 9, 1
 
\refb Wesson, P. S. 1982, Vistas Astron., 26, 225
 
\refb  Zeldovich, B. Ya. 1970, A\&A, 5, 84
  
\end{document}